\documentclass[11pt]{article}
\usepackage{graphicx}
\usepackage{graphicx,amsmath,bm,amssymb,color,theorem,algorithmicx,algorithm} 
\usepackage{epsfig}
\usepackage{epstopdf}
\newcommand{\bx}{\boldsymbol{x}}
\newcommand{\ba}{\boldsymbol{a}}
\newcommand{\bmu}{\boldsymbol{\mu}}

\newtheorem{Lemma}{Lemma}
\newtheorem{Corollary}{Corollary}

\newtheorem{Theorem}{Theorem}

\newcommand{\va}{\mbox{${\bf a}$}}

\newcommand{\vx}{\mbox{${\bf x}$}}

\newcommand{\vzero}{\mbox{${\bf 0}$}}
\newcommand{\vone}{\mbox{${\bf 1}$}}

\newcommand{\mI}{\hbox{{\bf I}}}
\newcommand{\mJ}{\mbox{${\mathbf J}$}}

\newcommand{\mW}{\hbox{{\bf W}}}
\newcommand{\mWa}{\mbox{${\bf W_{\ga}}$}}

\newcommand{\ga}{\alpha}

\newcommand{\gD}{\Delta}

\def\bm#1{\mbox{\boldmath $#1$}}

\newcommand{\vgp}{\mbox{$\bm \pi$}}

\newtheorem{question}{Question}[section]
\newtheorem{coro}{Corollary}[section]

\newcommand{\beq}{\begin{equation}}
\newcommand{\eeq}{\end{equation}}
\newcommand{\bea}{\begin{array}}
\newcommand{\ena}{\end{array}}
\newcommand{\bds}{\begin {itemize}}
\newcommand{\eds}{\end {itemize}}
\newcommand{\bdf}{\begin{definition}}
\newcommand{\blm}{\begin{lemma}}
\newcommand{\edf}{\end{definition}}
\newcommand{\elm}{\end{lemma}}
\newcommand{\bthm}{\begin{theorem}}
\newcommand{\ethm}{\end{theorem}}
\newcommand{\bprp}{\begin{prop}}
\newcommand{\eprp}{\end{prop}}
\newcommand{\bcl}{\begin{claim}}
\newcommand{\ecl}{\end{claim}}
\newcommand{\bcr}{\begin{coro}}
\newcommand{\ecr}{\end{coro}}
\newcommand{\bquest}{\begin{question}}
\newcommand{\equest}{\end{question}}

\newcommand{\rarrow}{{\rightarrow}}

\begin{document}
\title{The impact of random actions on opinion dynamics}

\author{Amir Leshem\thanks{Corresponding Author: Amir Leshem, Faculty of Engineering, Bar-Ilan University, Ramat-Gan, 52900, Israel. Phone: +972-35317409. email: leshema@ biu.ac.il.} and Anna Scaglione\thanks{Dept. of EE Arizona State University.}}

\maketitle
\begin{abstract}
Opinion dynamics have fascinated researchers for centuries. The ability of societies to learn as well as the emergence of irrational {\it herding} are equally evident. The simplest example is that of agents that have to determine a binary action, under peer pressure coming from the decisions observed.   By modifying several popular models for opinion dynamics so that agents internalize
actions rather than smooth estimates of what other people think, we are able to prove that almost surely the actions  final outcome remains random, even though actions can be consensual or polarized depending on the model. This is a theoretical confirmation that the mechanism that leads to the emergence of irrational herding behavior lies in the loss of nuanced information regarding the privately held beliefs behind the individuals decisions.
\end{abstract}




\maketitle

Mathematical models of herding phenomena prove the old saying that “actions speak louder than words”. A classical result shows that sequential Bayesian learners give rise to dysfunctional information cascades because of the overwhelming effect of the actions that are observed. In this work we demonstrate that a variety of popular opinion dynamics models, modified so that the agents internalize the effect of random actions rather than smooth estimates their peers beliefs,  create herding, polarized groups or do not converge to a fixed belief. Because of the generality of our analysis, our work cements the notion that actions are poor representations of private information and lead to unpredictable social herding phenomena.

There is a vast literature on the subject of {\it opinion dynamics}, aimed at modeling how beliefs propagate across a social fabric.
In the late 1700s \cite{condorcet} first noted that averaging individuals' beliefs would, by the law of large numbers, lead to a {\it noise free} average opinion and precise estimate in the limit \cite{condorcet}, a concept later popularized broadly in the early 1900's \cite{fgalton}. The models used to derive these insights are very similar to the Bayesian multi-agent models that are pervasive in signal processing \cite{Vikram14,krishnamurthy2014tutorial,Sayed14}. While in signal processing decision criteria are often based on minimizing the expected Bayesian risk,
in economics, the decision model of {\it rational agents} has been long modeled as maximizing the expected utility in the Bayesian sense. Furthermore, a substantial body of the economics literature assumes that agents reveal imperfectly their private beliefs through their utility maximizing action. The seminal paper by Bikhchandani, Hirshleifer and Welch (BHW) \cite{bhw92} (see also \cite{Lohmann94_2})  highlighted how pathologies, called {\it informational cascades}, seem to develop when agents make a zero-one decisions (for example, zero is ``bad'' and one is ``good''), instead of averaging continuous values as in \cite{fgalton}.  The BHW paper highlights the significant impact of processing discrete actions, as opposed to continuous belief values, on the final behavior of the group. This impact leads to what is often referred to as  {\it herding behavior} \cite{chamley2004rational}.    Other  opinion diffusions models and interactions of agents in social networks are reviewed in e.g. \cite{jackson,kleinberg,acemoglu,castellano2009,krishnamurthy2013social,friedkin2006structural}.

In the last forty years, a flurry of models for opinion dynamics has emerged in the statistical physics literature \cite{castellano2009}, inspired by ideas aimed at predicting macroscopic behavior that emerges from simple random particles interactions. Among them the {\it voter model}, introduced in \cite{clifford1973model}, is one of the simplest non-ergodic opinion diffusion models that leads to herding. This more closely resemble the zero-one interactions in \cite{bhw92} since the agents have two discrete opinions and copy at random one of their neighbors. This model was later mapped onto that of random walkers that coalesce upon encounter by Liggett \cite{liggett1985particle,holley1975ergodic} which made it analytically tractable. An interesting variant is the one introduced by Mobilia in \cite{mobilia2003does}
where, using a mean-field approach, the author studied the effect of having {\it stubborn} agents in the network that, by not changing their opinion, attempt to influence the final equilibrium point. In recent work \cite{yildiz2010voter,yildiz2011discrete,yildiz2013binary} re-examined this question and analytically characterized its behavior by generalizing the approach introduced by Liggett and providing a solution for the strategic placement of stubborn agents in the voter model. Stubborn agents were also considered by Jia et al. \cite{jia2015}

Sznajd and coauthors in \cite{sznajd2000opinion} considered a different discrete opinion dynamic in which it takes
two agents to convince their neighbors of the correctness of their opinion. By contrast, Galam in \cite{galam2002minority}, proposed a variant in which the opinions of the agents switched to the prevalent opinion (a {\it majority rule}) in their neighborhood. Since Sznajd dynamics are a special case of the
general sequential probabilistic model (GPM) in \cite{galam2005local}, only two different phases are possible: either consensus, or coexistence of the two opinions in equal proportions. Interestingly, in the presence of stubborn agents with asynchronous updates, opinion fluctuations exist and convergence is not certain \cite{acemouglu2013opinion}.

Continuous belief models and updates, even in the presence of non-linearities (e.g. bounded confidence) are also a popular way to model social dynamics of opinions.
A popular form is the  {\it Hegselmann-Krause} (HK) model (or the {\it bounded confidence} (BC) model) \cite{Hegselmann2002, Blondel2009}, where opinions are represented by a real number and are updated synchronously, by averaging all the agents' opinions that differ by less than a confidence level $\epsilon$ (set to 1 in  \cite{Blondel2009}). As long as the network is connected, the agents reach a consensus \cite{Tsitsiklis1984} with the average belief in the network.
A randomized variant of the HK was introduced by DeGroot in \cite{DeGroot} and studied in e.g. \cite{Deffuant, Weisbuch01interactingagents,soc-weisbuch04,li2013consensus} in which two randomly chosen neighbors meet and adjust their opinions only when their opinion distance is below a threshold. These models lead either to consensus or polarization, but opinions do converge.

 Note that in the economics literature, due to the Bayesian update rule, the agents have a continuous belief even if their action and decision variables are discrete (the probability mass function of the decision variable given the private information and history of the actions that were observed). Generalizing the analysis in \cite{bhw92} to arbitrary social graphs is both intractable and perhaps futile, given the ample evidence that social agents are not rational \cite{kahneman2003maps}.

 In statistical physics the acronym CODA (continuous opinions discrete actions) was coined much later in \cite{martins2008continuous,martins2009opinion}. This class of models follows the general idea from economics of assuming that the agents internalize an estimate of the beliefs of the others obtained by observing their actions. But contrary to the BHW model, CODA models postulate simple mechanisms of opinion contagion, without offering significant detail on the interpretation of the models and providing mostly numerical characterization of their asymptotic behavior. The strong guarantees and mathematical characterizations of the limiting behavior is completely known only for the simple chain of sequential interactions treated in \cite{bhw92}.

Our main results is to offer strong theoretical guarantees that the same emergent behavior of random herding in the BHW model holds more broadly. Our work supports the argument that it is the discrete and random nature of what the agents internalize about the beliefs held by their neighbors that produces the formation of random herds, i.e. herds whose final decision outcome that cannot be predicted with certainty and is, potentially, nefariously wrong.


\section{ Consensus under the social pressure of actions}\label{sec:consensus}

We assume each agent has only two options to chose from, and we label them choice $0$ and choice $1$. Assume these choices are made by flipping a coin that is biased by a belief variable $x_n(t)$, generating an action $a_n(t)\in\{0,1\}$. For each $n$, $a_n(t)$ is a Bernoulli random variable with probability $x_n(t)$.

To capture the agents' tendency to conform to social norms, we assume that the agents perturb their belief $x_n(t)$ by moving towards the weighted mean of their neighbors actions.
The agents update their belief as a convex combination of the frequency of action $1$ over their neighborhood and their prior belief, i.e.:
\begin{equation}
\begin{array}{ll}
x_n(t+1)	&=(1-\alpha)x_n(t)+\ga\sum_{k=1}^N w_{nk} a_k(t) \label{belief-update}
\end{array}
\end{equation}
where $w_{nk}$ is a weight function corresponding to the relative weight that agent $n$ gives to the action of agent $k$  if $(n,k)$ is and edge and zero otherwise. We also require that $\sum_{k} w_{n,k}=1$. We require that the matrix
$\mW$ defined by:
will be the transition matrix of an irreducible Markov chain but we do not require $w_{n,n}=0$ since it might be that agent $n$ is influenced by its own action (even though it is random and determined by its internal belief $x_n(t)$. Basic results from spectral graph theory show that this condition is equivalent to
having a network that has a single component, where we have weighted edges between agents $n,k$  whenever $w_{n,k}\neq0$.
Note that $x_n(t)$ is always an number in $[0,1]$ and assuming that not all $x_n(0)$ are identically $0$ or identically $1$,  we may assume without loss of generality that they are strictly between $0$ and $1$ (since this will happen with probability 1 in finite time by the averaging property).
%
Introducing the vectors $\bx(t)=(x_1(t),\ldots,x_N(t))^T$ and
$\ba(t)=(a_1(t),\ldots,a_N(t))$ we note that:
\begin{equation}\label{eq-netupdateEXP}
{\mathbb E}\{\ba(t)|\bx(t)\}=\bx(t),
\end{equation}
and
\beq
 \hbox{Cov}\{\ba(t)|\bx(t)\}=\hbox{diag}\left\{v_1(t),...,v_n(t) \right\}
\eeq
where $v_j(t)=x_j(t)(1-x_j(t))$ is the variance of agent $j$ at time $t$.
Furthermore,
\begin{align}\label{eq.P(a)}
Pr\left(\ba(t)=\ba|\bx(t)\right)&=\prod_{n=1}^Nx_n^{a_n}(t)(1-x_n(t))^{(1-a_n)}.
\end{align}
The value of $\ga$ determines the influence of the actions on the belief of each agent\footnote{We assume that $\ga$ is independent of $n$ for notational simplicity, but all arguments go through even when $\ga_n$ depends on the player.}.
Let us define also the matrix $\mW$ which has elements $w_{m,n}$ and define
\begin{equation}
\mW_{\ga}=(1-\ga)\mI+\ga\mW.
\label{eq.mW}
\end{equation}

We note that for all $0 \le \ga \le 1$ both $\mW$ and $\mWa$ are stochastic matrices with the same right and left eigenvectors, and spectral radius 1. Let $\vgp$ be the left eigenvector of $\mW$  corresponding to the largest eigenvalue $1$ (normalized such that $\vone^T \vgp=1$) and remember that the corresponding right eigenvector is $\vone$. By the Perron-Frobenius theorem  we have $\vgp>>\vzero$, i.e. all its components are strictly positive.
Denoting by
\begin{equation}\label{eq.J}
\mJ\triangleq {\vone} {\vgp}^T,
\end{equation}
By standard theory of stochastic matrices for all $\ga$ the spectral
radius of $\mWa-\mJ$
$0<\rho(\mWa-\mathbf{J})<1$ and $\mWa\mathbf{J}=\mathbf{J}$.
Introducing the vectors $\bx(t)=(x_1(t),\ldots,x_N(t))^T$ and
$\ba(t)=(a_1(t),\ldots,a_N(t))$, we can also note that \eqref{belief-update} can be written in vector form as
follows:
\begin{equation}
\label{eq-netupdate}
\bx(t+1)=(1-\alpha)\bx(t)+\ga\mW\ba(t)
\end{equation}
From \eqref{eq-netupdate} it is straightforward  to prove the following:
\begin{Lemma}\label{lem.ave}
In general ${\mathbb E}\{\bx(t+1)|\bx(t)\}=\mWa\bx(t)$ and ${\mathbb E}\{\mathbf{J}\bx(t+1)|\bx(t)\}=\mathbf{J}\bx(t)$.
\end{Lemma}
The proof of the lemma is provided in the supplementary material.
This lemma clarifies the connection with the classic opinion dynamics model introduced by DeGroot \cite{DeGroot}, which is identical to our model in expectation.
The weights in the matrix  $\mWa$ in the DeGroot model typically areinterpreted as being the {\it trust} individuals place on each other.
The average dynamics are also a classic example of the so called Average Consensus Gossiping (AGC) algorithms \cite{proceedings-paper}.
For a connected network is well known that these dynamics lead to consensus on the state.
Note however \eqref{eq-netupdate} is not deterministic and the fact that actions are discrete  leads to very different dynamics than AGC as the following lemma shows:
\begin{Lemma}
The network has only two stationary states:  $\bx(t)={\mathbf 1}$ and $\bx(t)={\mathbf 0}$.
\end{Lemma}
{\bf Proof:}
It is clear that in both
cases, $\bx(t)={\mathbf 0}$  or $\bx(t)={\mathbf 1}$,
if we evaluate \eqref{eq.P(a)} for $\bx(t)={\mathbf 0}$ $Pr\left(\ba(t)={\mathbf 0}|{\mathbf 0}\right)=1$ and $\bx(t)={\mathbf 1}$ $Pr\left(\ba(t)={\mathbf 1}|{\mathbf 1}\right)=1$. This implies \eqref{eq-netupdate} leads to a fixed point $\bx(t+1)=\bx(t)$. Any other belief level will place a non-zero probability in all $\ba(t+1) \in \{0,1\}^N$ therefore making it possible to deviate from the previous belief for every
$\alpha>0$.

Interestingly, the fact that there are stationary states, does not suffice to prove that the beliefs indeed converge to the stationary states. This convergence is the main result.
Denote the expected value of $\bx(t)$   by $\bmu_{\bx}(t)$:
\begin{equation}\label{mean-opinion}
\bmu_{\bx}(t)= \mathbb{E}\{\bx(t)\}.
\end{equation}
Let the network weighted sample average of the beliefs according to the stationary distribution $\vgp$ at time $t$ be:
\begin{equation}
q(t)=\vgp^T \bx(t).
\end{equation}
The next lemma leverages results in the AGC literature \cite{proceedings-paper}:
\begin{Lemma}\label{lemma-mean}
The expected value in \eqref{mean-opinion} is such that
\begin{equation}
\lim_{t\rightarrow +\infty} \bmu_{\bx}(t) ={\mathbb E}\{q(0)\} {\mathbf 1}.
\end{equation}
\end{Lemma}
The simple proof of this lemma is relegated to the supplementary material for brevity.
While this property is not surprising, what is interesting is the following result:
\begin{Theorem}\label{main-lemma}
The mechanism in \eqref{belief-update} leads to herding, i.e.
$$\forall n,~~~~~Pr\left(\lim_{t\rightarrow \infty} x_n(t)\in \{0,1\}\right)=1.$$
Moreover, the limit is identical to all agents, i.e., either all agents end up with belief $0$ or all of them end up with belief $1$.
\end{Theorem}
The first part of the proof of this lemma is similar to that used in \cite{jadbabaie2012non} albeit for a different opinion diffusion model  to show that the process converges to a random variable with probability 1. However, in contrast to \cite{jadbabaie2012non}, our model leads to herding while their model leads almost surely to consensus on the true value of the parameter. The second part resorts to the Lesbegue dominated convergence theorem to prove that the limiting random variable must be equal to either $0$ or $1$:\\
{\bf Proof:}~
We know that the only stationary belief levels are $\bx=\vzero$ or $\bx=\vone$, since for every other vector there is a positive probability that the average action will cause a deviation up to $\ga$, however, in contrast to the voter's model this is insufficient, since when the initial state has all beliefs strictly between $0$ and $1$, at any given stage the beliefs will be in the open interval $(0,1)$ since they are convex combinations of the current belief with a number in $[0,1]$ and convergence is not achieved in finite time.
Since $\mW$ is a stochastic matrix. Let $q(t)=\vgp^T\vx(t)$ be the average belief level in the network according to the stationary distribution of $\mWa$. We will show that $q(t)$ is a martingale process with respect to the beliefs.
To that end we observe that:
\begin{equation}
q(t+1)=\vgp^T \vx(t+1)
		=\vgp^T \left[(1-\ga)\vx(t)+(\ga\mW)\va(t)\right].
\end{equation}
where $\va(t)$ is the action vector at times $t$. Computing the expectation of both sides conditioned on $\vx(t)$ we observe that:
\beq
\label{cond_qone}
\mathbb{E}\left\{q(t+1)| \vx(t)\right\}=\vgp^T \mWa\vx(t)=\vgp^T \vx(t)=q(t),
\eeq
since $\vgp$ is a left eigenvector of $\mWa$ with an eigenvalue equal to 1 and by the Perron-Frobenius theorem  all its elements are strictly positive, since the network is connected (which is equivalent to the Markov chain defined by $\mW$ is irreducible.
This implies that $q(t)$ is a martingale with respect to the sigma-algebra determined by the sequence of beliefs $\vx(t)$. By definition $q(t)$ is a bounded sequence, as the weighted mean of a vector with elements between 0 and 1.
By the martingale convergence theorem the sequence $q(t)$ must converge to a random variable $q_{\infty}$ with probability 1. We need to show that $q_{\infty}$ is almost surely either $0$ or $1$,

Consider now for $t>1$ the sequence $\gD q(t)=q(t)-q(t-1)$, i.e. the martingale difference sequence.
For each $t$ by the martingale property,
\beq
E\left\{\gD q(t)\right\}= 0.
\eeq
Since $\vgp$ is a probability vector and for all $n,t$  $0 \le x_n(t) \le 1$ we have that for all $n,t$ $|q(t)|<1$ as well as $|\gD q(t)|<1$. Therefore, by the almost sure convergence of $q(t)$
\beq
\lim_{t \rarrow \infty} {\rm var}(\gD q(t))=0,  \ \ \hbox{a.s.}
\eeq
which implies that also
\beq
\lim_{t \rarrow \infty} {\rm var}(\gD q(t)|q(t-1))=0,  \ \ \hbox{a.s.}
\eeq
Similarly $\gD q(t) <1$ and converges almost surely to $0$.  Hence by the Lesbegue dominated convergence theorem $\gD q(t)$ converges to $0$ in the mean square sense.
If we express explicitly the conditional variance of  $\gD q(t)$
\beq
\bea{lcl}
{\rm var}(\gD q(t)|q(t-1))&=&E\left\{\gD q^2(t)|q(t-1)\right\}\\
				&=&\alpha^2\sum_{n=1}^N \pi_n^2 {\rm var}(a_n(t)) \\
                          &=&\alpha^2\sum_{n=1}^N \pi_n^2 x_n(t)(1-x_n(t)).
\ena
\eeq
Since for all $n$, $\pi_n>0$ the MS convergence implies that
\beq
\lim_{t \rarrow \infty} x_n(t)(1-x_n(t))=0~~~~\forall~n.
\eeq
Simple probabilistic computation shows that either for all $n$
\beq
\lim_{t \rarrow \infty} x_n(t)=0~~\mbox{or}~~
\lim_{t \rarrow \infty} x_n(t)=1.
\eeq
In fact, if for some $j,n$ which are connected, $x_n(t)$ converges to $1$ while $x_j(t)$ converges to $0$, since the actions are independent, there will be with probability $1$ infinitely many deviations of at least $\min\left\{|W_{n,k}| : w_{n,k}\neq 0 \right\}$ of both $x_n(t)$ and $x_j(t)$, contradicting the convergence of both $x_n(t)$ and $x_j(t)$.

This implies  that herding must be achieved with probability 1, which ends the proof of the main theorem.

It should be noted that while the limit variables $x_n(\infty)$ have values $0,1$, each initial condition has a different probability of converging to $0$ or $1$, which implies that what outcomes prevails, wrong or right, is random and therefore social pressure can lead to umpredictable collective behavior, in spite of the initial information the agents have available to forge their opinion. We would like to determine its distribution given an initial value or a distribution of initial values.
The next corollary will show that $x_n(\infty)$ is a binomial random variable with mean $q(0)$. This implies that the variance of  $x_n(\infty)$ is $q(0)\left(1-q(0)\right)$.
In fact, the combination of the lemma above and the main theorem leads to the following corollary:
\begin{Corollary}
Let $\vx(0)$ be the agents' initial belief level vector. The probability that the agents herd to action $a_n(t)=1$ is $q(0)=\vone^T\vx(0)$, i.e.
$\lim_{t\rightarrow +\infty} Pr(x_n(t)=1)=q(0)$.
\end{Corollary}
{\bf Proof:}
Since, as $t\rightarrow +\infty$, the network can only be in one of the two fixed points $\bx(t)={\mathbf 1}$
or  $\bx(t)={\mathbf 0}$, the probability that it will endup on one of these two states is, as indicated in Lemma \ref{lemma-mean}, equal to the average initial belief.

Interestingly, no matter what the initial state is, we cannot predict the outcome of the herding for sure. No matter how close the population belief is to $0$ or $1$ initially, it is always possible that they will herd towards the opposite action.
If it is indeed true that the social pressure mechanism is based on internalizing the actions of one's peers, we cannot predict accurately the outcome of the social dynamics. Applying this to market behavior, where people favor one product over the other, this indicates a quality brand that has high probability of capturing a large percentage of the market if individuals were to choose in isolation, may actually fail and other firms with worse products  have a fighting chance to capture it. Furthermore, we can see that even in large markets driven only by the consumers repeated actions, a {\it winner takes all} cascade eventually happens, and one product wins the majority of the market share irrespective of the initial individual evidence to the quality of the product. This may not always be the case however. In the next section we generalize our analysis first to random interactions and show the same type of herding phenomenon happens as well in that context. We also show that when social pressure is mitigated by mistrust for what deviates excessively from ones prior belief, then the society may split in multiple herds.

\subsection{Randomized interactions}\label{sec.randomdyn}
A simple variation of our model is one that captures random interactions among the agents.
In this case \eqref{belief-update} becomes
\begin{equation}
\begin{array}{ll}
x_n(t+1)	&=(1-\alpha)x_n(t)+\ga\sum_{k=1}^N w_{nk}(t) a_k(t) \label{belief-update-random}
\end{array}
\end{equation}
where the weights $w_{nk}(t)$ are non zero only for those individuals who observe and are affected by their peers actions in that particular epoch. We assume $w_{nk}(t)$ are random processes independent from the actions the nodes performed, which continue to be modeled as in the previous section. Correspondingly, we can define the random matrices $\mW(t)$ and $\mW_{\alpha}(t)$ as before. Interestingly this variation of the model hardly changes the result and random herding ensues also in this case almost surely.
Let us now assume that these random matrices are ergodic processes and denote by $\mW=\mathbb{E}\{\mW(t)\}$. If all nodes will speak infinitely often with peers and the average
$\mW=\mathbb{E}\{\mW(t)\}$ is a stochastic matrix with eigenvalue 1 with multiplicity 1, then we can easily generalize the proof of our main theorem, using the same definition of $\vgp$ as before:
\begin{Corollary}\label{main-lemma-for-randomW}
The statement of Theorem \ref{main-lemma} holds unchanged if
the interactions are random and $\mW=\mathbb{E}\{\mW(t)\}$.
\end{Corollary}
{\bf Proof:}~
In this case we define:
\beq
q(t)=\vgp^T\vx(t)
\eeq
In this case, like before:
\beq
\label{cond_qone_rW}
\mathbb{E}\left\{q(t+1)| \vx(t)\right\}=\vgp^T \mWa\vx(t)=\vgp^T \vx(t)=q(t),
\eeq
but now the average is also with respect  to the process $\mW(t)$. Nonetheless
everything else from this point on follows the steps of the proof of
Theorem \ref{main-lemma}.

\section{Bounded confidence models}\label{sec.bounded}

\subsection{Bounded confidence under the influence of actions}
An interesting generalization of our model is the update with {\it bounded-confidence},
where the $n$'th agent updates happen only where the observations are sufficiently close to
the agent own disposition.
A natural option is to analyze a model similar to the HK
model proposed in \cite{Hegselmann2002} for continuous opinion dynamics. In the HK model, agents mix their belief only with agents whose belief sufficiently close to their own. In our case, agents update their belief only if the empirical distribution of the actions
of its neighbors is sufficiently close to $x_n(t)$. We introduce the function:
\beq
\rho(x)=\alpha xu(|x|\leq\tau)
\eeq
in which the confidence threshold $\tau<1$ and $u(x)$ is the Heaviside (i.e. the unit step) function. The agent's update with bounded confidence in its neighbors is modeled as:
\beq\label{eq.udpdate-BCm}
x_n(t+1)=x_n(t)+\rho\left({\sum_{k=1}^N} w_{nk}a_k(t)-x_n(t)\right).
\eeq
We note right away that $\rho(0-0)=\rho(1-1)=0,$
which implies that $\vx(t)=\vone$ or $\vx(t)=\vzero$ are, once again fixed points.
Let us denote by:
\begin{eqnarray}
P_n(\vx(t))=P\left(  \Big| {\sum_{k=1}^N} w_{nk}a_k(t)-x_n(t)\Big|>\tau \Big|\vx(t)\right)
\end{eqnarray}
Let:
\beq
\alpha_n(\vx(t))=\alpha (1-P_n(\vx(t))).
\eeq
We can define a matrix $\mW(\vx(t))$ in the same way as \eqref{eq.mW}, except that we replace $\alpha$ with a diagonal matrix with diagonal elements $\alpha_n(\vx(t))$.
It is straightforward to see that
$$\mathbb{E}\{\vx(t+1)|\vx(t)\}=\mW(\vx(t))\vx(t)$$
which are smooth
non-linear dynamics equivalent to the HK model.
In the action-based HK dynamics a range of stationary beliefs are possible and, also the beliefs may never converge even in distribution.
In fact, next we provide an example where the action-based HK model exhibits much more complex behavior than the polarization phenomenon observed in the classical HK model.
Let us illustrate this fact with an example. Consider a network with four nodes, with an initial beliefs vector $\vx=(0,0.45,0.55,1)^T$, a confidence threshold $\tau=0.25$ and a mixing matrix:
\begin{equation}\label{eq:4nHK}
\mW=
\left[
\begin{array}{cccc}
 0 & 1  & 0 &  0\\
 0.5 & 0  & 0.25  & 0.25\\
0.25  & 0.25  & 0  & 0.5\\
0 & 0  & 1 &  0
\end{array}
\right].
\end{equation}
In this case the dynamics of the four nodes are as follows. For the first node:
\begin{align}
x_1(t+1)&=\left\{
\begin{array}{ll}\!\!\!
x_1(t)\!+\!\alpha u(|1\!-\!x_1(t)|\leq \tau)(1\!-\!x_1(t)) & a_2(t)=1\\
\!\!\!x_1(t)\!-\!\alpha u(x_1(t)\leq \tau)x_1(t)  & a_2(t)=0
\end{array}
\right.\\
&=
\left\{
\begin{array}{ll}\!\!\!
x_1(t)=0 & a_2(t)=1\\
\!\!\!(1\!-\!\alpha)x_1(t)=0  & a_2(t)=0
\end{array}
\right.
\end{align}
and the reason why $x_1(t)=0$ is that we assumed that $x_1(0)=0$ and all later values remain unchanged.
Similarly, without giving a proof we can show that $x_4(t)=1$ (just by symmetry $1-x_4(t)$ behaves like
$x_1(t)$). For node number two, when update is activated, the node has a new state that is a convex combination of $x_2(t)$ with either $0.25$ or $0.5$. Therefore:
\begin{equation}
(1-\alpha)x_2(t)+\alpha ~0.25\leq x_2(t+1)\leq (1-\alpha)x_2(t)+\alpha ~0.5
\end{equation}
and, since the node starts at $x_2(0)=0.45$, the subsequent values will have to remain confined in the interval $(0.25,0.5)$. For similar reasons $x_3(t)$ will remain in the interval $(0.5,0.75)$.
However, they both will change randomly. In fact, $x_2(t)$ will change due to $a_3(t)$ and viceversa $x_3(t)$
will change based on $a_2(t)$. Therefore no convergence in distribution is attained
and the opinions will continue to fluctuate indefinitely in the intervals discussed above.
This is shown numerically in the next section in Fig. \ref{fig:HK4nodesim}.


\subsection{Reinforcement model with random pairwise interactions}
We propose an alternative model for the update. We assume that if agents $n$ observe at random one of its neighbors, say the $k$th neighbor, make the same decision, agent $n$ will adjust the belief as follows:
\beq
x_n(t+1)=x_n(t)+\alpha\delta[a_n(t)-a_k(t)](a_n(t)-x_n(t)),
\label{eq.reinf}
\eeq
where $\delta[x]$ is the Kronecker delta function.

We view this model as a mechanism that reinforces behavior, because it tends to decrease or increase the belief based on the frequency with which ones action is repeated in the neighborhood.
In this case the only fixed points for the dynamics $\vx$ have integer entries, but unlike either \eqref{cond_qone} or \eqref{cond_qone_rW}, in this case the network may never coalesce in taking a single action, but opposite decisions may persist as well.
The expectation of these dynamics given a random pair $(n,k)$ is:
\begin{align}
{\mathbb E}\{x_n(t+1)|\vx(t),(n,k)\}&=x_n(t)[1-\alpha(1-x_n(t))(1-x_k(t))]\nonumber\\
&\le x_n(t).
\end{align}
Since this holds for each random choice of $(n,k)$ we observe that from the previous equations, taking the expectation also over the random choices of $(n,k)$  it is clear that the state is a sub-martingale:
\beq
{\mathbb E}\{x_n(t+1)|\vx(t)\} \le x_n(t)
\eeq
Next we prove the following lemma:
\begin{Theorem}\label{th:reinforcement}
The mechanism in \eqref{eq.reinf}  leads to
$$\forall n,~~~~~Pr\left(\lim_{t\rightarrow \infty} x_n(t)\in \{0,1\}\right)=1.$$
but not necessarily herding, which in turn means society will be polarized in general.
\end{Theorem}
{\bf Proof}:~
The argument follows a similar line of reasoning as our previous theorem. Because of the dominated convergence theorem the state $x_n(t)$ must converge in the mean square sense. We can see then
that the sequence $y_{nk}(t)=\alpha\delta[a_n(t)-a_k(t)](a_n(t)-x_n(t))$ must go to zero in the mean
square sense, because each pair is selected at random infinitely often.
This can happen in only two cases, either the conditional variance of $a_n(t)-x_n(t)$ goes to zero
or $\delta[a_n(t)-a_k(t)]=0$ with probability one. If we assume the first is not true, then the second
implies that the independent actions of the random pair of agents are the same, which can only be
if they have equal probability $x_n(t)=x_k(t)\in\{0,1\}$; while this cannot happen in finite time
it can happen at the limit. If this does not happen then the variance of $a_n(t)$ in the limit must be zero
which again implies that in the limit the probability converges to either 1 or 0. Both lead to the same conclusion, which proves the theorem.

\section{Simulated experiments}
\subsection{ Consensus under the social pressure of actions}
\begin{figure}[ht]
\centering
	\includegraphics[width=0.8\textwidth]{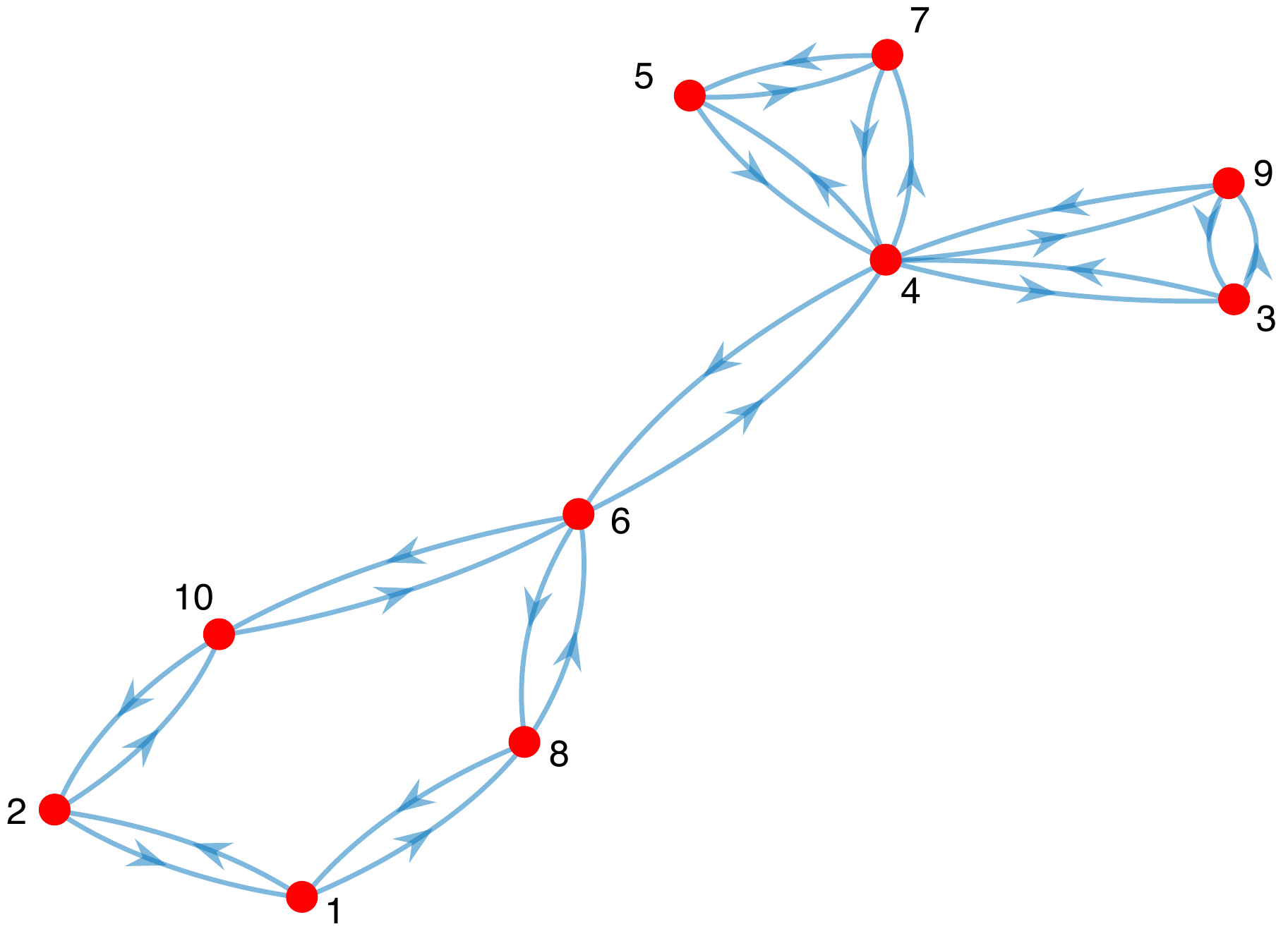}
\caption{The random graph used for simulations.}
\label{fig:graph}
\end{figure}

To demonstrate the herding and clustering phenomenon we proved to be true in Section \ref{sec:consensus}, we performed several Monte-Carlo trials. In the first experiments we randomly picked a random sample graph shown in Figure \ref{fig:graph}.

Then we picked randomly $5000$ initial vectors $\vx(0)$ whose entries are independent and drawn to have a mean $\forall~n~\mathbb{E}\{x_n(0)\}=p_0$ varying from $0.2$ to $0.8$ and chose uniform weights $w_{kn}=1/deg(k)$, where $deg(k)$ denotes the degree of node $k$ (the graph is undirected).

\begin{figure}[ht]
\centering
	\includegraphics[width=0.8\textwidth]{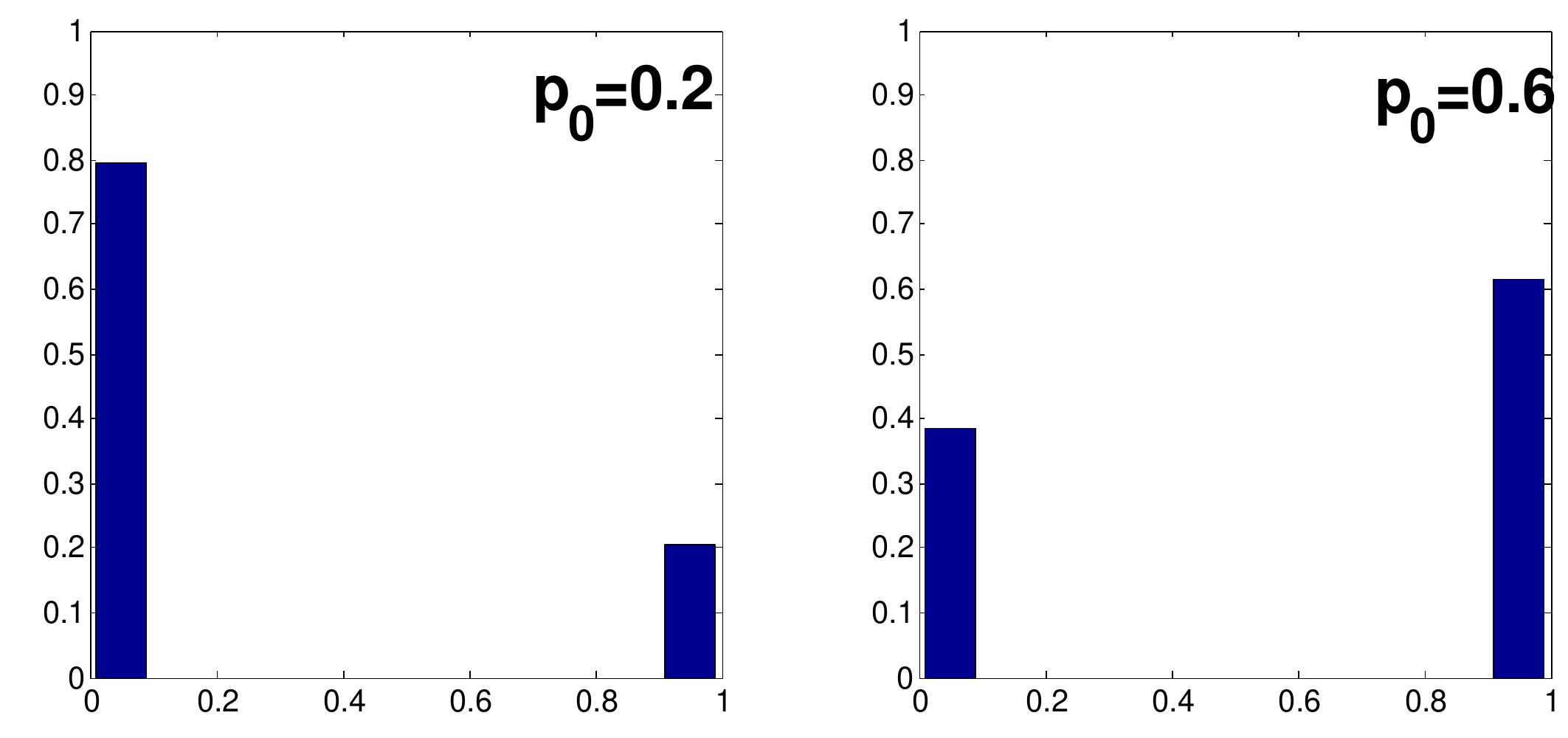}
\caption{Histograms of the final belief for $p_0=0.2$ and $p_0=0.6$.}
\label{fig:hist}
\end{figure}

\begin{figure}[ht]
\centering
	\includegraphics[width=0.8\textwidth]{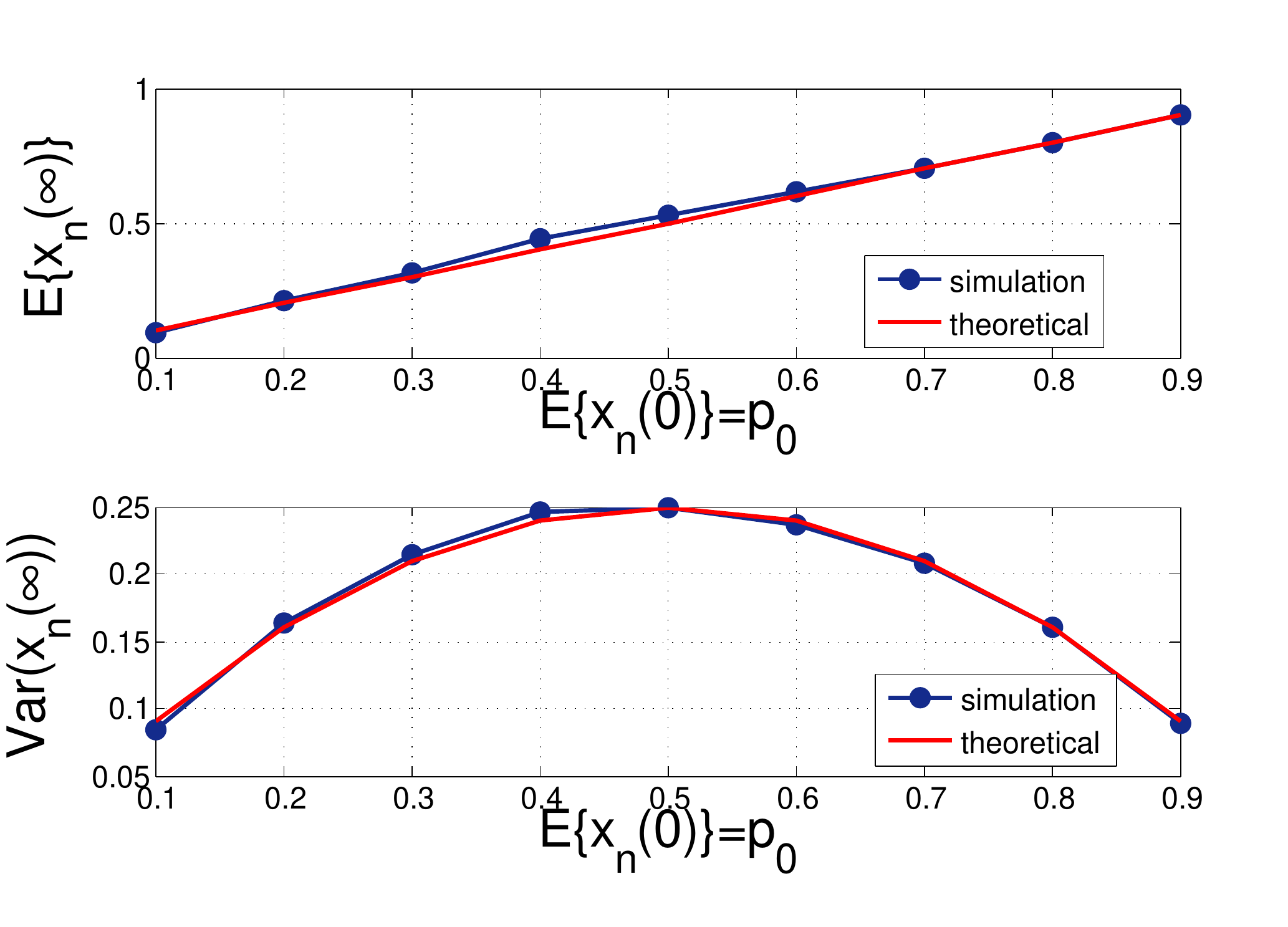}
\caption{Mean and variance of the limit point as a function of $p_0$.}
\label{fig:mean_variance}
\end{figure}

Figure \ref{fig:hist} shows the histogram of the limit for various random realizations of the process for initial belief levels of $p_0=0.2,0.6$. Figure \ref{fig:mean_variance} shows clearly that the actions are Bernoulli random variables whose parameter is correctly predicted by the theorem.

\subsection{Action-based HK model}
\begin{figure}[ht]
\centering
	\includegraphics[width=0.8\textwidth]{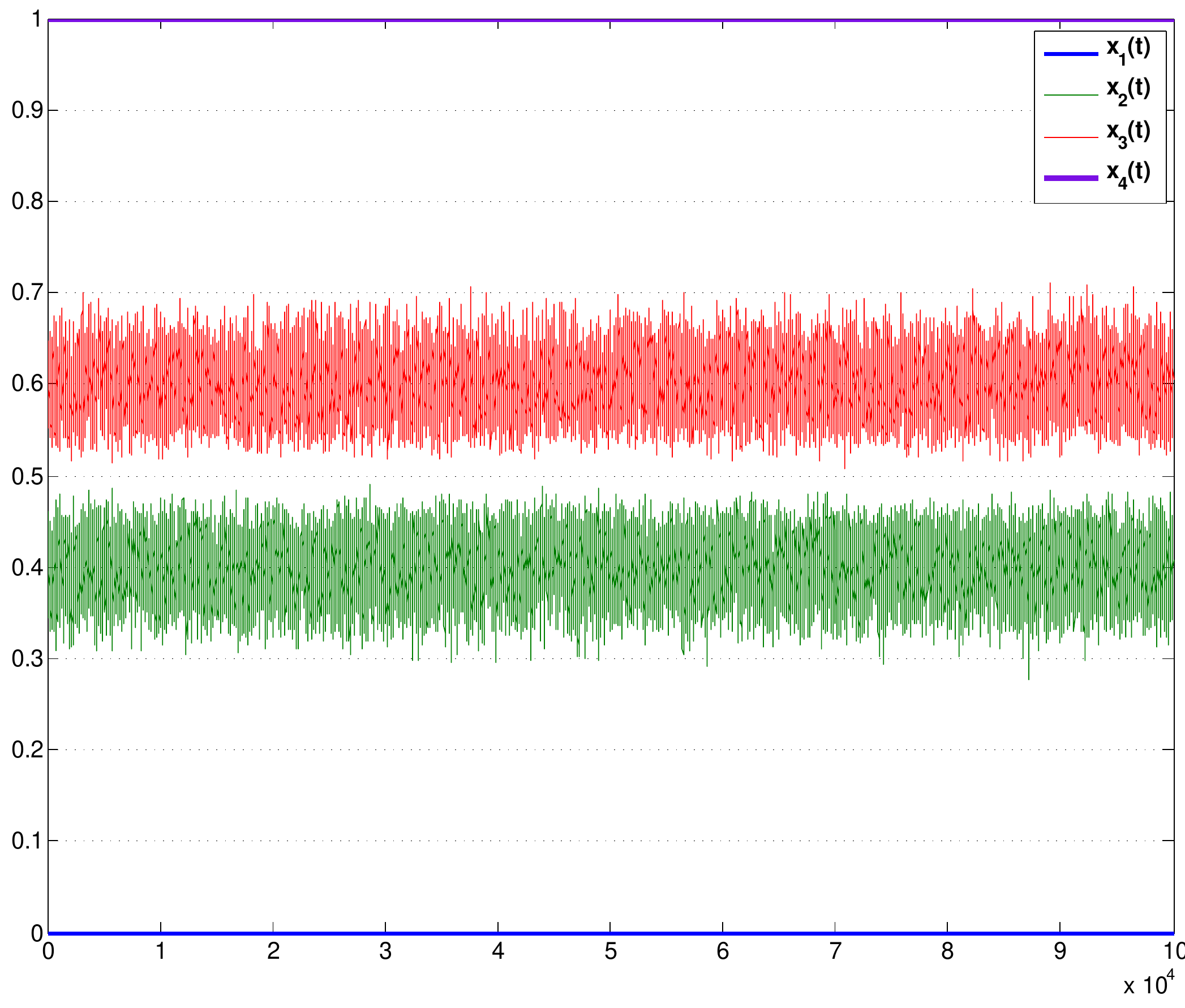}
\caption{The state of the four node network with $\mW$ in (\ref{eq:4nHK}).}
\label{fig:HK4nodesim}
\end{figure}

In Figure \ref{fig:HK4nodesim} we simulated the dynamics discussed in Section \ref{sec.bounded}
pertaining four nodes  with $\mW$ in (\ref{eq:4nHK}).
Note that there is no agreement but also no convergence of belief, in contrast to the classical HK model and the claims made in Section \ref{sec.bounded} are corroborated by the simulation results. Furthermore the random process of the opinions of the agents is very complex, as one can notice from the complex non linear dynamics of the mean. This shows that it is much harder to predict
a form of polarization because the random processes $x_2(t),x_3(t)$ are governed by non-linear stochastic dynamics that are Markovian but non homogeneous, whose trends are difficult to manipulate mathematically to obtain optimum forecasts. Either than resorting to the bounds we mentioned,
accurate predictions become intractable, even when the initial conditions are known.

\subsection{Reinforcement model}
\begin{figure}[ht]
\centering
	\includegraphics[width=0.8\textwidth]{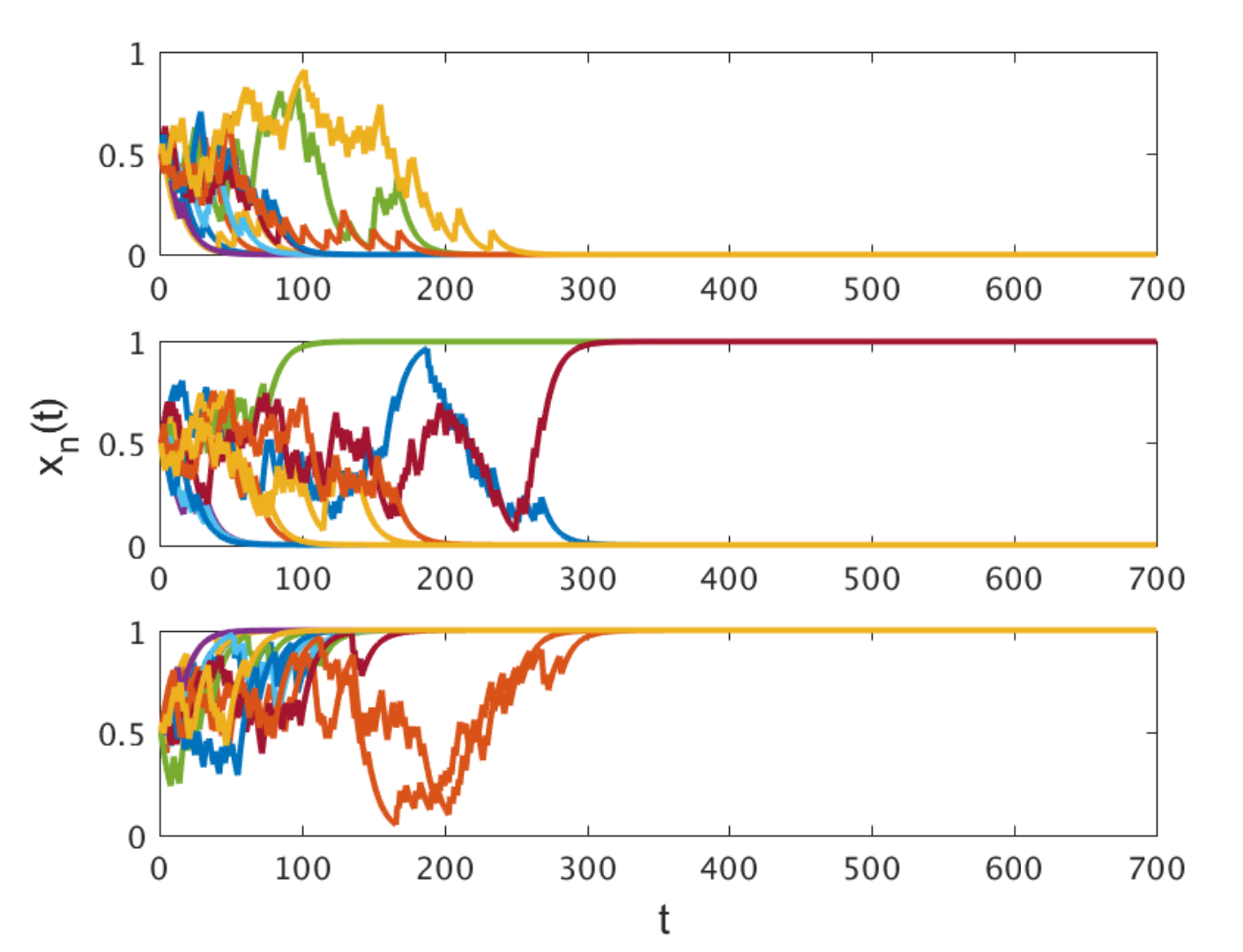}
\caption{Three experiments of the dynamics of the agents beliefs with same initial beliefs that converge to herding (top 0 and bottom 1) and polarization (middle).}
\label{fig:reinforcement}
\end{figure}

In Section \ref{sec.bounded} we introduced a new model in which nodes move their belief when their action is  identical to one of their neighbors, selected at random. In this case we showed that the network in Figure \ref{fig:graph}, with uniform initial belief equal to 0.5, converges to possibly multiple herds. This is illustrated in our three numerical simulation of the dynamics of the beliefs in  Figure \ref{fig:reinforcement} where we can clearly observe the emergence of herding towards belief 1 and belief 0 by all nodes in the top and bottom figures respectively, as well as the emergence of polarized groups, whose belief is 1 or 0,  in the middle plot. This corroborate the statement in Theorem \ref{th:reinforcement}.

\section{Conclusions}
In this paper we examined an opinion diffusion model in which agents update their belief based on their neighbors' empirical distribution of the actions. The use of the observed actions leads to herding
unlike continuous state updates when neither bounded confidence nor stubborn agents are present in the system showing that the trend observed in the BHW model extends to other settings.



%


\section{Appendix}
\subsection*{Proof of Lemma 3}
We now prove prove Lemma 3
Taking the expectation on both sides of \ref{eq-netupdate}, we obtain:
\begin{equation}
\begin{array}{lcl}
\bmu_{\bx}(t+1)&=&(1-\alpha){\mathbb E}\{\bx(t)\}+\ga\mW{\mathbb E}\{{\mathbb E}\{\ba(t)|\bx(t)\}\} \\
			&=&\mWa\bmu_{\bx}(t)=\mWa^t\bmu_{\bx}(0)
\end{array}
\end{equation}
which is the AGC update that converges to consensus in the limit. In fact
\beq
\lim_{t\rightarrow +\infty}\mW^t=\mathbf{J}
\eeq
and
\beq
\mathbf{J}\bmu_{\bx}(0)=\vone \vgp^T\bmu_{\bx}(0)
\eeq
which proves the claim, since ${\mathbb E}\{q(0)\} =\vgp^T\bmu_{\bx}(0)$.
In addition to the lemmas and theorems in the main paper, the following result holds
\begin{Lemma}\label{lem.ave1}
In general
\begin{equation}
{\mathbb E}\{\bx(t+1)|\bx(t)\}=\mWa\bx(t)
\end{equation}
and
\begin{equation}
{\mathbb E}\{\mathbf{J}\bx(t+1)|\bx(t)\}=\mathbf{J}\bx(t).
\end{equation}
\end{Lemma}
{\bf Proof:}
\begin{equation}
\begin{array}{lcl}
{\mathbb E}\{\bx(t+1)|\bx(t)\}&=&(1-\alpha)\bx(t)+\ga\mW {\mathbb E}\{\ba(t+1)|\bx(t)\}\\					
                              &=&(1-\alpha)\bx(t)+\ga\mW\bx(t)=\mWa\bx(t).
\end{array}
\end{equation}
The second claim follows easily since $\mathbf{J}\mWa=\mathbf{J}$.
This shows that in expectation the belief follows the same dynamics as the DeGroot model \cite{DeGroot}.
\subsection*{Simulations of herding under random pairwise interactions}
\begin{figure}[ht]
\centering
	\includegraphics[width=0.8\textwidth]{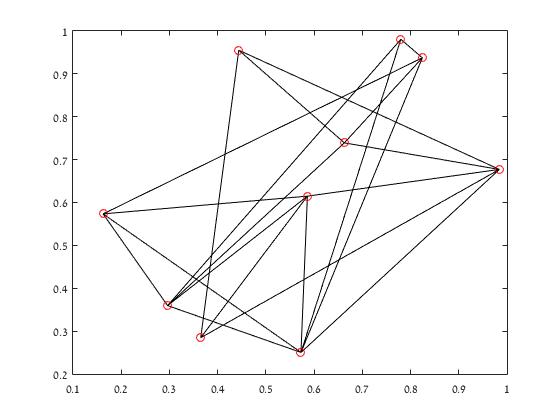}
\caption{10 nodes networks used for the asynchronous dynamics.}
\label{fig:graph_gossip}
\end{figure}

\begin{figure}[ht]
\centering
	\includegraphics[width=0.8\textwidth]{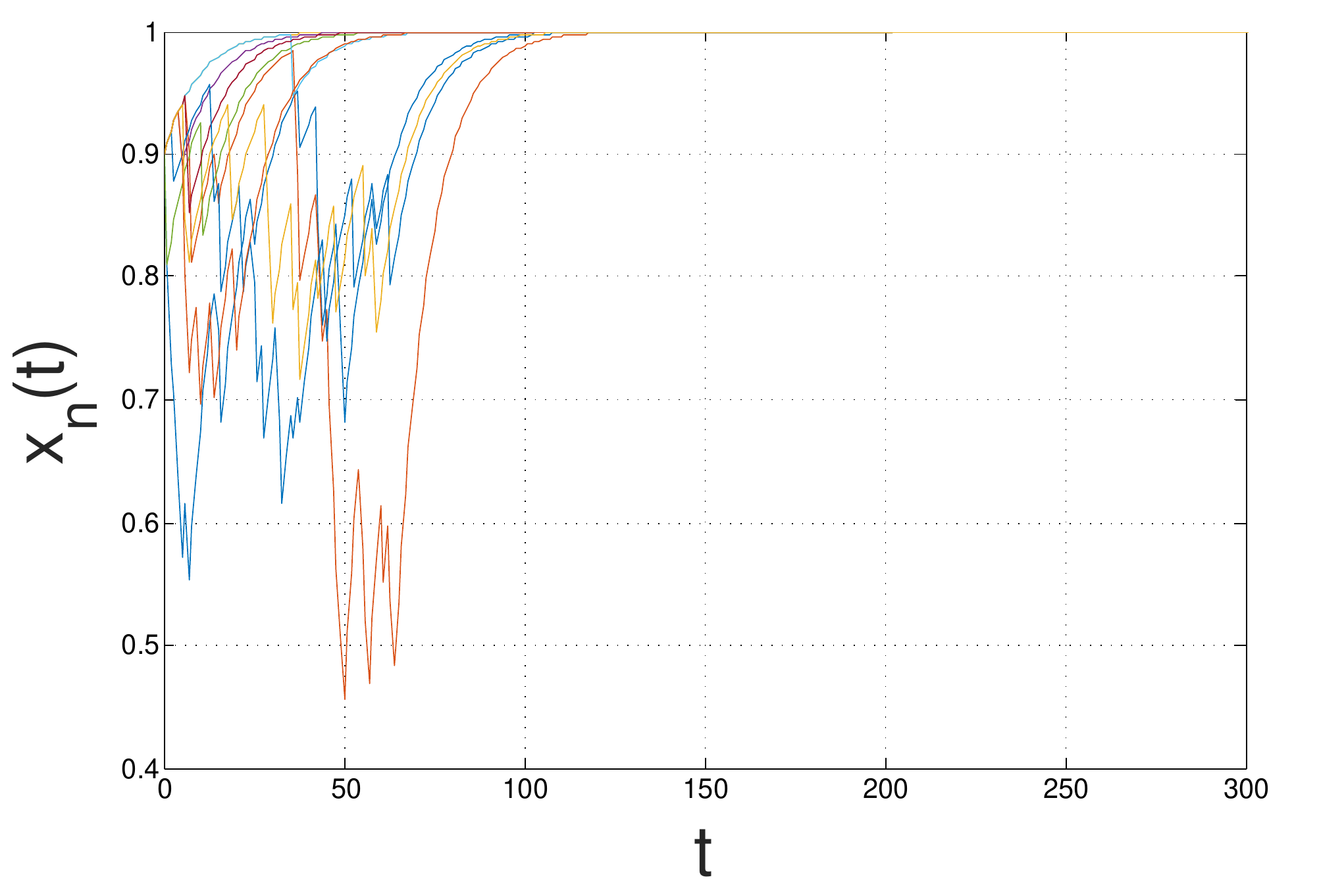}\\
	\includegraphics[width=0.8\textwidth]{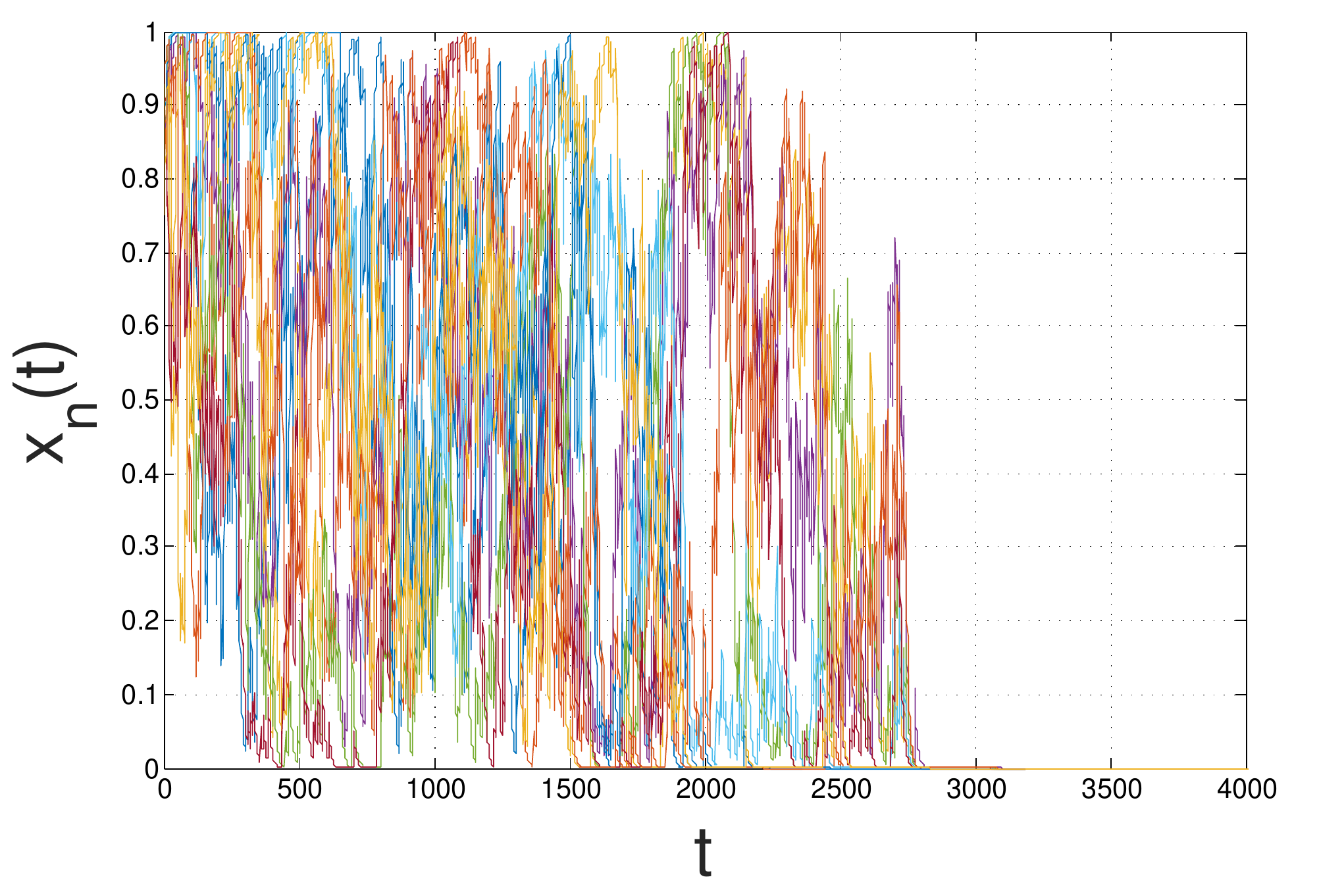}
\caption{Opinion dynamics herding to $1$ (top) and to $0$ (bottom), $x_n(0)=0.9,~\forall n$.}
\label{fig:convergeto0}
\end{figure}

We also tested the dynamics where each time two random nodes average their actions, discussed in the Randomized interactions section, and use the averaged action to update their mutual beliefs. Figure \ref{fig:graph_gossip} describes the random graph used. In this case we describe the course of two dynamics of the population, both where all nodes had an initial belief of $0.9$, i.e. all nodes were very inclined to take action $1$. The top Figure \ref{fig:convergeto0} presents a dynamics converging to $1$. We present the lowest belief, the highest belief and the mean belief. As can be seen when the dynamics converge to $1$ this happens quite rapidly, because of the initial state of all nodes.
The bottom of Figure \ref{fig:convergeto0} presents another instance of the dynamics, where initial beliefs were identical to the previous case, but the limiting value of the herding was $0$. With the initial conditions defined in the simulation, this occurs only $10\%$ of the simulations. Still this is surprising, as it would be expected that the initial belief will have significant impact on the results of the herding. As predicted by the main theorem, the direction of the herding is unexpected, and governed by the random actions of the agents.
\newpage


\end{document}